\documentclass[aps,10pt,prb,twocolumn,showpacs,groupedaddress]
{revtex4-1}  
\usepackage{graphicx}  
\usepackage{dcolumn}   
\usepackage{amssymb}   
\usepackage{array}
\usepackage{amsmath}
\usepackage{tabularx}
\usepackage{natbib}
\usepackage{notes2bib}
\usepackage{epstopdf}
\begin{document}

\title{Diamond Magnetometry of Superconducting Thin Films}

\author{A. Waxman$^{1}$}
\email {amirwaxm@bgu.ac.il}
\author{Y. Schlussel$^{1}$}
\author{D. Groswasser$^{1}$}
\author{V.M. Acosta$^2$}
\author{L.-S. Bouchard $^3$}
\author{D. Budker $^4$}
\author{R. Folman$^1$}
\affiliation{$^1$ Department of Physics, Ben-Gurion University of the Negev, Be'er Sheva 84105, Israel\\
$^2$ Hewlett-Packard Laboratories, 1501 Page Mill Rd., Palo Alto, California 94304, USA\\
$^3$ Department of Chemistry and Biochemistry, Bioengineering and California NanoSystems Institute, University of California, Los Angeles, California 90095, USA\\
$^4$ Department of Physics, University of California at Berkeley, Berkeley, CA 94720-7300, USA}

\date{\today}

\begin{abstract}
In recent years diamond magnetometers based on the nitrogen-vacancy (NV) center have been of considerable interest for applications at the nanoscale.  An interesting application which is well suited for NV centers is the study of nanoscale magnetic phenomena in superconducting materials.  We employ NV centers to interrogate magnetic properties of a thin-layer yttrium barium copper oxide (YBCO) superconductor. Using fluorescence-microscopy methods and samples integrated with an NV sensor on a microchip, we measure the temperature of phase transition in the layer to be $70.0(2)$ K, and the penetration field of vortices to be $46(4)$ G. We observe pinning of the vortices in the layer at $65$ K, and estimate their density after cooling the sample in a $\sim 10$ G field to be $0.45(1)$ $\mu$m$^{-2}$. These measurements are done with a 10 nm thick NV layer, so that high spatial resolution may be enabled in the future. Based on these results, we anticipate that this magnetometer could be useful for imaging  the structure and dynamics of vortices. As an outlook, we present a fabrication method for a superconductor chip designed for this purpose.
\end{abstract}

\maketitle

\section{Introduction}

The study of magnetic phenomena in superconductors and the search for superconductors possessing high critical temperatures ($T_c$)~\cite{PIC2006,ZAK1998,REY2007} require adequate measurement techniques.   Direct signatures of superconductivity (diamagnetism and vanishing resistivity) may be complemented by measurements of one or more properties such as local density of states, nuclear magnetism, or heat capacity. Of particular interest is the study of magnetic vortices in type-II superconductors, which is the focus of this work.  Type-II superconductors exhibit phase transitions at two critical magnetic field values, with the lower denoted $H_{c1}$ and the higher, $H_{c2}$. For magnetic fields $H<H_{c1}$, these superconductors exhibit the Meissner effect, whereby magnetic flux is expelled from the interior. In the mixed state, $H_{c1}<H<H_{c2}$, magnetic flux can penetrate through the cores of superconducting vortices. Each vortex, which carries a single quantum of magnetic flux, consists of a core, where superconductivity is suppressed within a radius $\sim\xi$ (the coherence length), surrounded by circulating supercurrents, which persist over a length scale $\sim\lambda$ (the penetration depth).

The structure and magnetic properties of vortices are of interest in the study of pnictides, which feature irregular arrays of vortices~\cite{bib:eskildsen}, and in the search for multi-component superconductors that are predicted to contain vortices with fractional multiples of the flux quantum~\cite{bib:chibotaru,bib:geurts}.   The motion of vortices is also of interest, because it leads to energy loss that degrades the performance of almost all superconducting devices.   Vortex motion also determines the critical current density of superconductors, see for example Ref. \cite{bib:marchevsky}, and can serve as a model for condensed-matter flow~\cite{bib:paltiel}.  Reduction of vortex motion has been achieved through the use and engineering of pinning centers~\cite{bib:beleggia,bib:loudon}.

The physics of pinning centers has been explored mainly through macroscopic measurements of properties such as electrotransport and bulk magnetization. These integrated-response techniques conceal the details of the microscopic properties of vortex pinning.  Thus, efforts have been devoted toward developing real-space imaging methods for direct visualization of vortex patterns.  Magnetic imaging enables obtaining accurate values of the penetration depth, which reports on the number density of electrons involved in superconductivity, the nature of the superconducting state,~\cite{bib:moshchalkov} and the types of vortex interactions which can occur~\cite{bib:chaves}.    Aside from capturing vortex structure, techniques that feature video frame rate may enable studies of vortex dynamics~\cite{bib:loudon,bib:kramer}.

Several methods have been developed for vortex visualization~\cite{bib:bending} such as magnetic-resonance force microscopy (MRFM)~\cite{bib:luan_moler}, scanning magnetic probes~\cite{bib:moser,bib:kirtley,bib:oral} (including scanning Hall-probe microscopy~\cite{bib:menghini}), Lorentz microscopy~\cite{bib:harada}, magneto-optical imaging systems (polarized-light microscopy)~\cite{bib:koblischka,bib:polyanskii}, and transmission electron microscopy (TEM)~\cite{bib:loudon13}.  TEM offers high spatial (better than $20$ nm) and temporal resolution. A tilted sample features vortices that penetrate normally to the surface of the film to provide a component of the B-field normal to the electron beam, causing the electrons to be deflected by the Lorentz force and appear as black-white features in an out-of-focus image~\cite{bib:harada,bib:loudon12,bib:loudon13}. However, TEM has been limited, so far, to low external fields ($<30$ G)~\cite{bib:loudon13}, whereas many type-II superconductors such as Nb$_3$Sn and NbTi possess upper critical fields that are much higher. Furthermore, the destructive nature of the measurement, which causes rapid damage to the sample, prohibits studies over long periods of time.

The MRFM method can achieve spatial resolution
which is similar to that of TEM \cite{RUG2004}. However, this measurement technique perturbs the magnetism of the sample to be measured, as it is based on getting the sample's magnetization to oscillate at the cantilever frequency. In addition, the MRFM sensitivity is severely compromised at higher sample temperature making it a less-than-ideal technique for the imaging of high Tc superconductors.

The quest for non-destructive and non-perturbative
methods has led to the development of magneto-optical
imaging systems. These systems take advantage of
magneto-optical materials that change the polarization
of light in proportion to the surface magnetic field. In recent years this technique has become a leading method in vortex imaging, giving rise to sub-micron spatial resolution and a~$\sim10\,\rm\mu T$ magnetic field resolution \cite{GOA2003,VES2007,GOL2009}. Nevertheless, the imaging of a single vortex remains a challenge
for this technique, as it is hard to keep a sub-micron gap, necessary for this kind of imaging, between the magneto-optical layer and the superconducting sample. In addition, the sensitivity reported to date~($\sim10\,\rm\mu T$) may be insufficient for accurate measurements of the vortex properties.

Magnetometry with nitrogen-vacancy (NV) centers in diamond is a good candidate technique for non-destructive and non-perturbative sensing with high spatial and temporal resolution across a wide range of external magnetic fields and sample temperatures. High spatial resolution can be achieved in a scanning-probe configuration~\cite{BAL2008,MAL2012,GRI2013} or through the use of sub-diffraction imaging methods, such as  stimulated emission depletion (STED) and ground-state depletion (GSD), where resolutions down to several nanometers are  possible~\cite{RIT2010}. Such a spatial resolution may allow visualization of the vortex core or imaging individual superconducting nano clusters.
The sensitivity of NV centers in diamond, which was investigated at the single NV center level~\cite{BAL2008,TAY2008,MAZ2008} and for ensembles~\cite{STE2010,PHA2011}, is such that NV-diamond sensors are capable of detecting  electron spins~\cite{bib:rugarPRB} and nuclear spins~\cite{bib:rugar2013science} located externally to the diamond. Finally, by depositing or growing the superconductor film directly on the diamond or by attaching two highly polished surfaces, it is expected that the sensor-sample gap would be in the nanometer range.

In a previous study, detection of the Meissner effect in a type-II superconductor with NV centers was demonstrated ~\cite{BOU2011} by sensing the fringe field of a macroscopic sample. The approach investigated in the present work is to generate a $10$ nm thin layer of NV centers within $\sim 25\ $nm of the surface of the diamond and detect the shift of their magnetic resonance using a focused laser beam.  The beam waist is approximately $1$ $\mu$m in diameter, leading to detection volumes on the order of $10^{-20}$ m$^3$.  A thin layer with a small sensor-sample gap enables high spatial resolution.

Recently, vortices in permalloy thin films were imaged using a single NV center~\cite{RON2012,TET2012,RON2013}. We choose instead to use an ensemble of NV centers, which will ultimately enable us to image the magnetic field in a larger area in a single shot (i.e. no scanning) and with a higher sensitivity ($\delta B\propto 1/\sqrt{N}$). In addition, ensembles enable the simultaneous measurement of all three components of the magnetic-field vector. Due to these properties ensemble magnetometers have recently been utilized for live cell imaging \cite{LES2013}.

In this study we utilize an NV sensor to characterize a thin layer ($300$ nm) of type-II superconducting yttrium barium copper oxide (YBCO) material. Using this sensor, we observe the Meissner effect and characterize the superconducting phase transition. We also demonstrate the ability to monitor the penetration field of vortices, locally. Finally, as an outlook, we describe a new superconducting device containing a micro-patterned superconducting layer which is designed to test the resolution of the sensor and its ability to map single vortices. We discuss the implications for imaging individual vortices or arrays of vortices.

\section{Experimental technique}
\label{Experimental technique}
\begin{figure*}[t]
\centering
  \begin{tabular}{@{}cc @{}}
    \includegraphics[width=8.6cm]{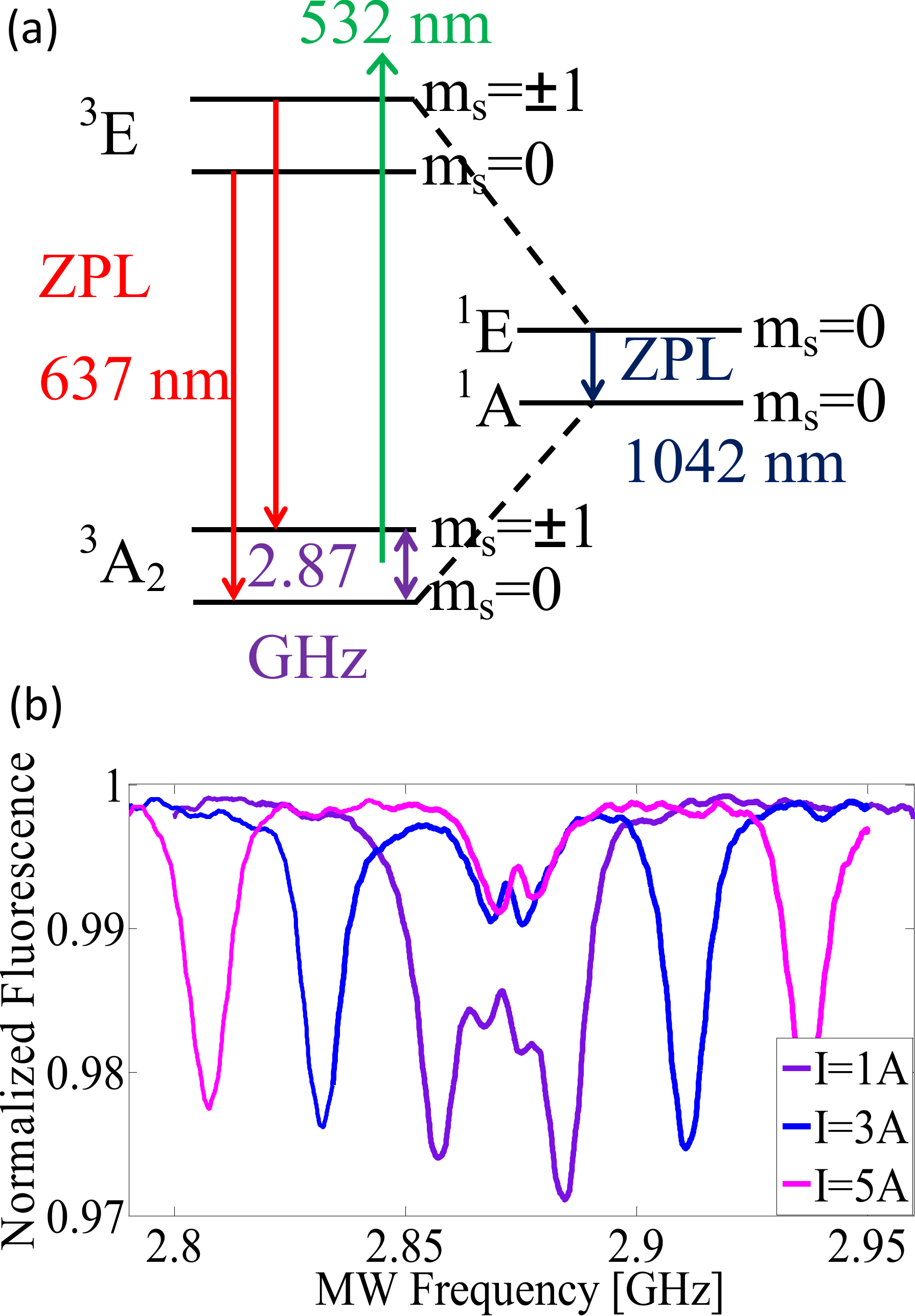}
    \includegraphics[width=8.6cm]{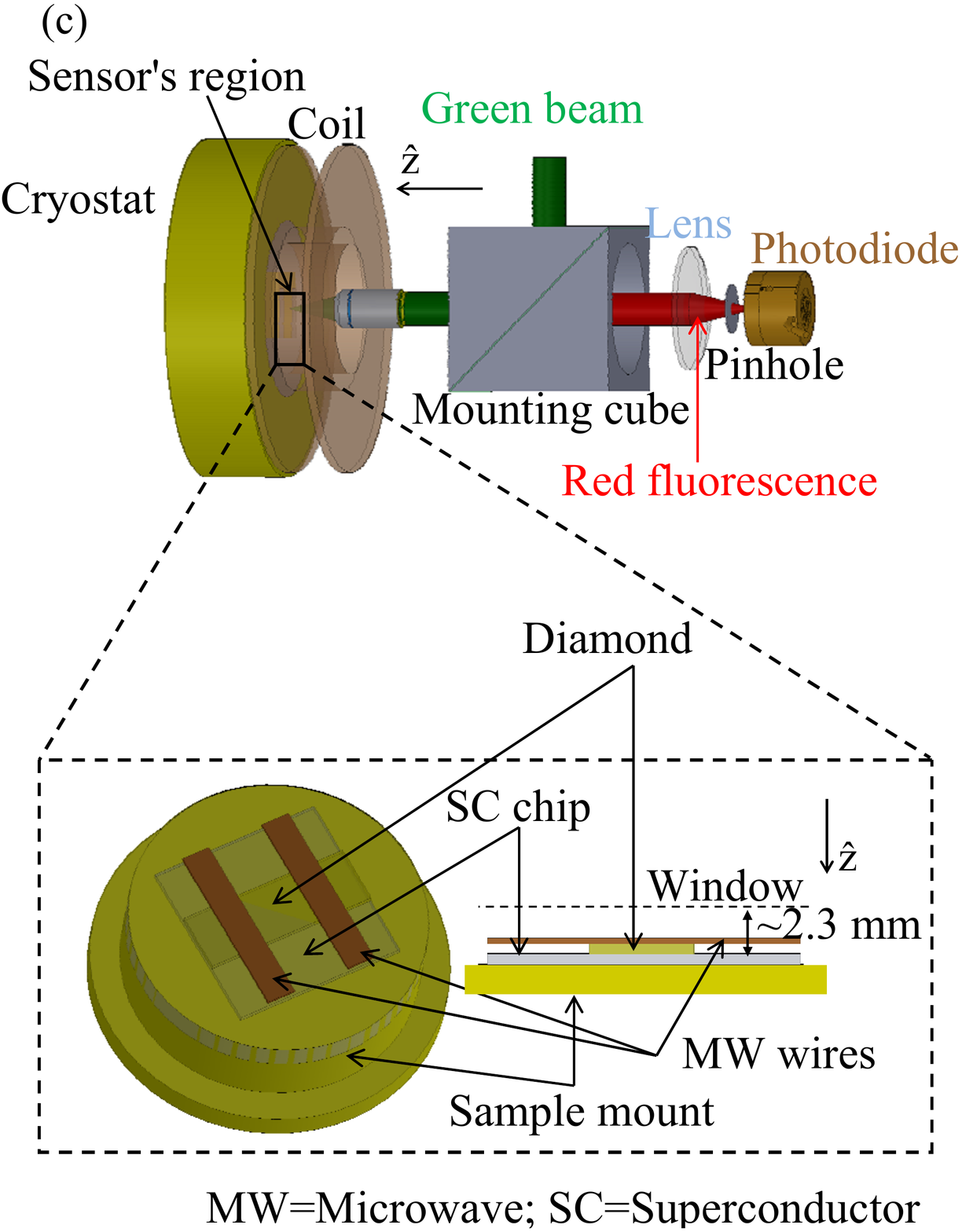}
\end{tabular}
\caption{(Color online) (a) Level diagram of the NV center. $^3A_2$ and $^3E$ are the ground and excited triplet states respectively. $^1A_1$ and $^1E$ are the intermediate singlet states involved in the optical-pumping process. The various spin levels are denoted by black lines. The radiative (non radiative) transitions are denoted by solid (dashed) arrows. (b) The ODMR spectrum taken at different external magnetic fields applied perpendicular to the surface of the diamond. The field was generated with a single coil driven by the currents indicated in the legend. The diamond used in the experiment is cut along a ($110$) plane, and the green light beam is perpendicular to its surface. (c) The experimental setup. The sample area is shown enlarged at the bottom. The distance between the NV sensor and the outer side of the cryostat window is $\sim 2.3$ mm. Such a distance enables us to focus the green light onto the sensor, as the maximal working distance of the objective is $4$ mm. (SC, YBCO superconductor; MW, microwave)}
\label{fig:basic_nv}
\end{figure*}

In this work we measure the magnetic field using the optically detected magnetic resonance (ODMR) method. We collect the red and near-infrared fluorescence emitted by NV centers when they are illuminated with green light. While the green light pumps the population to the $m_s=0$ spin level, microwaves scanned around $2.87$ GHz repopulate the $m_s=\pm1$ levels [see level diagram in Fig. \ref{fig:basic_nv}(a)~\cite{ACO2010singlet}]. This results in a reduction in the emitted fluorescence, as the centers in the $m_s=\pm1$ sublevels have higher probability of intersystem crossing into singlet states, where they are temporarily removed from the absorption-emission cycle (for more information regarding ODMR see Ref.\ \cite{ACO2013}).

Figure \ref{fig:basic_nv}(b) shows typical ODMR spectra measured with our diamond, for different values of applied external magnetic field.
There are four possible alignments of the NV axis in a diamond crystal. The diamond surface is polished along a (110) plane, meaning that two of the four possible alignments of the NV axes are at $\arccos(\sqrt{2/3})\approx 35^\circ$ with respect to the normal to the crystal plane (the out-of-plane axes) and the other two are at $90^\circ$ (in-plane axes). The laser beam is normal to the diamond surface (i.e., light propagates along the $\hat{z}$ direction). Applying a magnetic field along $\hat{z}$ using a coil results in a Zeeman splitting of the $\pm 1$ ground-state magnetic sublevels associated with the out-of-plane axes. This splitting is readily observed in the ODMR spectrum.  On the plot, we indicate the current running through the coil (1 A corresponds to approximately $4.5$ G). Since the NV centers are primarily sensitive to the component of the field along the NV axis, the resonances corresponding to the $90^\circ$ alignments are not significantly shifted, in contrast to those  at $35^\circ$ that split already at relatively low fields. The resonance positions are given by $D \pm g\mu_B B \cos35^\circ$, where $D\approx 2.87\ $GHz is the axial zero-field splitting, $g=2.003$ is the Land\'{e} factor, and $\mu_B= 1.40$~MHz/G is the Bohr magneton. From this frequency splitting we can extract the magnetic field in the~$\hat{z}$ direction. Note that in this specific case we do not exploit the vectorial nature attributed to NV ensemble
magnetometry as we foreknow the field of vortices will
be in the same direction of the applied field~($\hat{z}$). In the general case of an arbitrary field, which is not discussed in this paper, the spectrum will be split according to the four orientations, and then both the amplitude and the direction of the field can be extracted.

The experimental setup is shown in Fig. \ref{fig:basic_nv}(c). Confocal microscopy is performed by excitation with green light (532 nm) supplied by a diode pumped solid-state laser. The output beam is expanded to a diameter of $\sim1$ cm, larger than the diameter of the objective lens. This enables us to scan the beam across the sample, by moving the objective, without affecting the beam direction.
The objective (Olympus, Pro Plan) has a numerical aperture N.A.=$0.65$ and a maximal working distance of $4$ mm. The laser light power at the NV location was measured to be $10$ mW, while the diameter of the beam at the focal point is $\sim 1\,\mu$m assuming a diffraction limited spot.

The fluorescence emitted by the NV centers is collected with the same objective used to focus the incident green light, and is transmitted through two dichroic mirrors, both mounted in cubes to facilitate optical alignment (note we have used two mirrors here instead of the one usually used to ensure the complete filtering of the green light reflections). These cubes are mounted on a $3$D translation stage which is controlled by actuators. A lens is used to focus the fluorescence on a high-sensitivity photodiode (NewFocus 2151). To achieve higher spatial resolution it is possible to place a pinhole at the focal point of the lens (the detector is slightly moved back-
wards in this case) to ensure that only fluorescence from the focal plane of the objective is collected.

The superconductor sample is mounted on the copper cold finger of a cryostat (Janis model ST-$500$) using a vacuum compatible varnish (VGE 7031 from LakeShore). To promote good thermal contact, we use a Teflon piece [not shown in Fig. \ref {fig:basic_nv}(c)] which is held by screws and presses the superconductor sample against the cold finger.  We use a YBCO superconductor layer (obtained from Theva) with a thickness of $300$ nm, which was deposited on a MgO substrate ($5\times 5 \times 0.5$ mm$^3$). The $c$-axis of the superconductor is perpendicular to the plane of the film, so that the CuO$_2$ planes are parallel to the surface.   Transmission of the microwaves to the diamond is achieved with two copper strips placed along either side of the diamond on top of the MgO substrate and the YBCO layer [see Fig. \ref{fig:basic_nv}(c)].

To ensure we measure only the magnetic fields associated with the superconductor layer, compensation coils are employed. These coils enable the zeroing of the Earth magnetic field as well as other static-field sources.

As noted, the sensor used in this work is a diamond plate which contains a layer of NV centers near the surface. Since the field in the center of a magnetic vortex decays at short distances approximately as $B_0e^{-z/\lambda}$  ($B_0$ being the field at the surface, $z$ the distance from the surface, and $\lambda$ the field penetration depth; for YBCO $\lambda\approx 150$ nm), there is a gradient of the field over the sensor that leads to line broadening \cite{*[{To be precise, above a layer with a thickness $d$, the field does not decay exponentially, but rather falls off as $1/(z+\lambda_{eff})^2$} as mentioned by ] [{. Here $\lambda_{eff}=\lambda \coth[\lambda/2d]$ is the effective penetration depth. However, close to the surface ($0<z<1\,\mu$m) which is the case discussed here, the exponential approximation is good enough.}] CHA1992}. Thus a thin sensor layer is required to minimize broadening thereby increasing sensitivity. An optimal thickness must be found as too thin a layer reduces sensitivity due to the small number of NV centers. In any case, the thickness cannot be higher than the required spatial resolution for the same reason that the NV layer needs to be close to the diamond surface and consequently to the sample.

\begin{figure}[h]
\includegraphics[width=8.6cm]{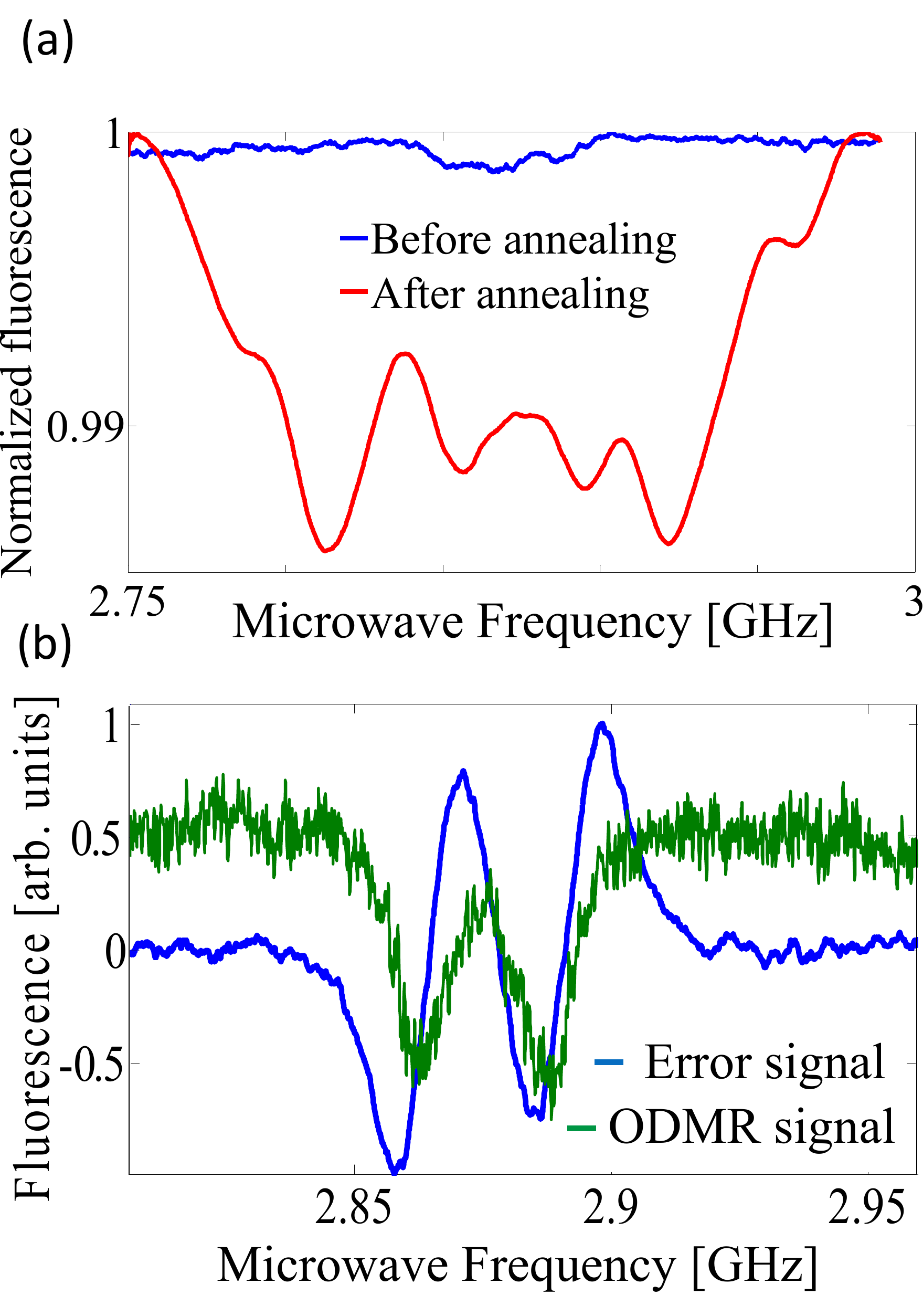}
\caption{Signal enhancement of the diamond sensor. (a) A comparison between the ODMR spectrum, before (blue line) and after (red line) the sample was annealed in oxygen. The contrast is enhanced due to conversion of NV$^0$ centers to NV$^-$. (b) An error signal (blue line) derived from the fluorescence signal (green line). The signal-to-noise ratio (SNR) is highly enhanced.}
\label{sample_prep}
\end{figure}

To meet these requirements, a diamond with a thickness of $80\,\mu$m produced by Element Six was implanted with N$^{+}$ ions at Core Technologies. The ion beam energy and the irradiation dose were $10$ keV and $10^{13}$ cm$^{-2}$, respectively.  Monte-Carlo simulations using the software of Ref.\ \cite{SRIM} indicate that the resulting layer of the implanted nitrogen atoms is located between $z\approx 15$ nm and $z\approx 25$ nm, where $z=0$ is at the diamond surface. To generate NV centers, the diamond was annealed in an inert atmosphere (Ar) at $800^\circ$C. Following Ref. \cite{FU2010}, we also annealed the diamond in an oxygen atmosphere ($60/40\%$ of Ar/O$_2$) at $400^\circ$C to enhance conversion of neutrally charged centers (NV$^0$) into negatively charged ones (NV$^-$).  The NV$^-$ centers are used for magnetometry, whereas the NV$^0$ centers produce undesirable background fluorescence. The effect of the extra annealing is seen in Fig.\ \ref{sample_prep}(a) as a much higher signal contrast of the annealed sample.  We estimate that one to five percent \cite{RAB2006} of the implanted ions formed NV$^-$ centers, and therefore the density of NV$^-$ centers is $\sim 1-5 \times 10^{11}$ cm$^{-2}$. Focusing the laser beam to a size of $1\,\mu$m$^2$ should yield $\sim 1,000-5,000$ NV$^-$ defects within the sensing volume.

The intrinsic sensitivity of the sensor depends mainly on the diamond characteristics. Using a lock-in amplifier, we can suppress noise from external sources and improve the signal-to-noise ratio, as shown in Fig. \ref{sample_prep}(b). The lock-in technique used here involves frequency modulating the scanning microwaves with a modulation depth of $6$ MHz and  a frequency of $500$ Hz (this signal also serves as the lock-in amplifier reference). As a result we observe an error signal at the lock-in output, which is effectively a derivative of the ODMR spectral profile. This technique amplifies the signal and reduces high-frequency noise.

The sensitivity of the sensor is calculated using
\begin{equation}
\delta B=\frac{\delta S}{dS/dB},
\label{eq:sensitivity}
\end{equation}
where $\delta S$ is the standard deviation of the signal, and $dS/dB$ is the slope. In the ideal case,
\begin{equation}
dS/dB\approx \frac{Rg\mu_B}{\Delta\nu},
\label{eq:slope}
\end{equation}
where $R$ is the contrast of the signal and $\Delta\nu\approx5$ MHz is the resonance width. Introducing the relevant numbers from our experiment (including $\delta S$ which is directly measured from the signal), we find $\delta B\approx 2\,\mu$T/$\sqrt{\mbox{Hz}}$. This sensitivity may be further improved through the diamond sensor optimization. The above sensitivity satisfies the demands of the current experiment.

\section{Experimental results}

\begin{figure*}
\centering
\includegraphics[scale=0.6]{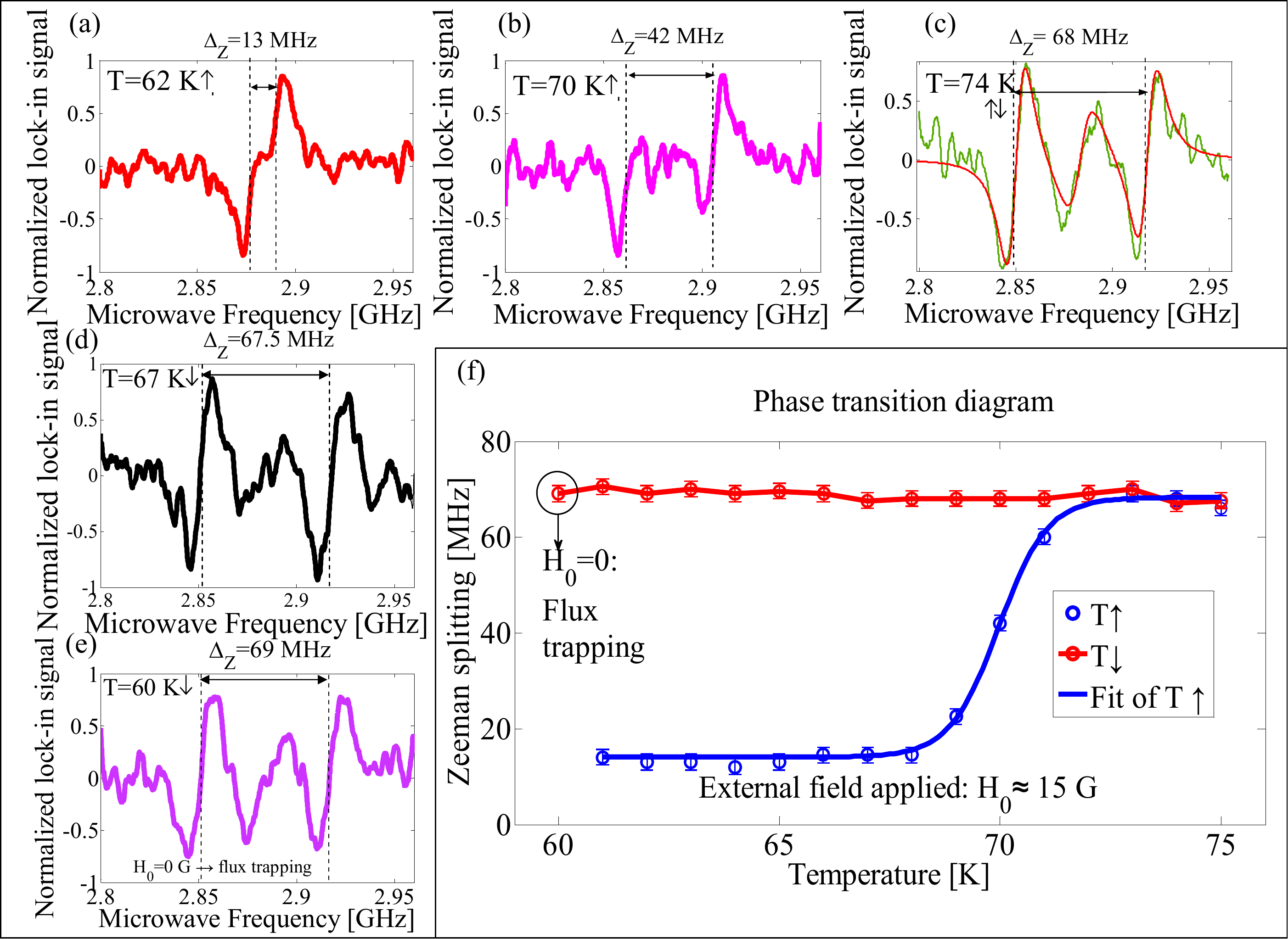}
\caption{(a-e) Several ODMR signals taken at different temperatures with an applied external magnetic field of 15 G.
The dashed black lines denote the Zeeman splitting, for each case. The up/down arrows indicate if the temperature was raised or lowered. (a) An ODMR signal taken below $T_c$, at $62$ K. The system is in the Meissner state, and the external field does not penetrate the superconductor layer. (b) The signal at the phase transition. Here we see a partial penetration of the field (a Zeeman splitting of $42$ MHz corresponds to a $\sim10$ G field in the $\hat{z}$ direction). (c) This signal, at $74$ K, indicates that the system is no longer in the superconducting phase. In this graph we also demonstrate the fit to a derivative of a Lorentzian function, used to determine the zero crossing of the error signal. We use this fit on all ODMR signals to find the value of the Zeeman splitting. (d) Signal, after taking the temperature down again, without turning the external field off. Since the critical field at $T_c$ is zero, vortices penetrate the layer, leading to the average magnetic field which we measure. (e) As detailed in the text, at $60$ K the magnetic field was turned off. Defects in the superconductor layer lead to flux pinning, evidenced by the field measured by the NV centers. (f) The phase-transition curve of the superconductor layer. Plotted is the Zeeman splitting between the $0\rightarrow1$ and the $0\rightarrow-1$ resonances of the $35^\circ$ NV-axis alignments. The blue data points belong to the ascending temperature sequence, while the red ones---to the  descending sequence. The external field during the measurements was $\sim15$ G excluding the last measurement (at $T=60$ K), where the field was turned off. We fit the blue data points to a sigmoid function (blue line), and extract a critical temperature of $T_c=70.0\pm0.2$ K. The red line is to guide the eye.}
\label{fig:transition}
\end{figure*}

After cooling to $\sim 60$ K, the sample was heated in steps of $1$ K while recording the ODMR spectra at each step. During these measurements, an external field of $\sim 15$ G was applied along $\hat{z}$, which would produce an ODMR splitting of $\sim69$ MHz [$=\rm 5.6\,MHz/G\times15\,G\times\cos(35^\circ)$] in the absence of a superconducting sample. However, none of the ODMR spectra below $T_c$ showed any Zeeman splitting due to this field [see Fig.~\ref{fig:transition}(a)]. A small splitting of 13 MHz is observed, but this splitting is also observed in the absence of external fields and can be attributed to an intrinsic non-axial strain field within the sensor \cite{JAN2012}. At $T=70$ K, Fig.\ \ref{fig:transition}(b), we observe a larger splitting, 42 MHz, which is still less than the expected Zeeman splitting in the absence of the superconductor. This indicates the onset of the superconducting phase transition. At $T$=74 K [see Fig.~\ref{fig:transition}(c)] we finally measured a splitting in the ODMR signal which corresponds to the external field. We then reduced the temperature below $T_c$, namely to $T$=67 K [see Fig.~\ref{fig:transition}(d)] and found that the Zeeman splitting remains largely unchanged from the value above $T_c$. The hysteresis can be attributed to the vortices created as we have entered the mixed state of the superconductor (for H-T diagrams of superconductors see Ref.~\cite{SCH1997}). Upon reducing the temperature further to $T$=60 K, we turned off the applied field, Fig.~\ref{fig:transition}(e). Strikingly, the ODMR spectrum retains the splitting observed for $T>T_c$ when the field was applied. This is the signature of flux trapping, whereby vortices remained pinned to defects in the superconductor even in the absence of an applied field. Figure \ref{fig:transition}(f) summarizes the sequence described above, and presents the Zeeman splitting variation as a function of temperature for both ascending and descending sequences.

Since the signal obtained with the lock-in amplifier is the derivative of the ODMR spectrum, we have analyzed the signal at each temperature by fitting a sum of $N$ Lorentzian derivatives ($N$ being the number of resonances in the spectrum),
\begin{equation}
F(\omega)=\sum_{i=1}^N  A_i \frac{ -2\gamma_i  (\omega-\omega_i)}{[(\omega-\omega_i)^2+\gamma_i^2]^2},
\label{eq:lorentz}
\end{equation}
where $\gamma_i$ is the linewidth of the $i$-th resonance and $\omega_i$ is its center, or the zero crossing of the error signal. $A_i$ are the resonance amplitudes. As a measure of the magnetic field, we use the frequency separation between the magnetic resonances, $0\rightarrow1$ and $0\rightarrow-1$, of the $35^\circ$ orientations. The error in the determination of this separation is calculated from the fit.

The accuracy of the fitting depends on the phase of the lock-in signal. We have adjusted the phase to be a multiple of $\pi/2$, meaning that on one channel the error signal is maximized whereas on the other channel the original ODMR signal is observed. This method yields reliable zero-crossing values when fitting.

%

Fitting the blue dots of the experimental data to a sigmoid function~\cite{BOU2011}
\begin{equation}
\Delta_Z(T)=\frac{a}{1+\exp[-(T-T_c)/\Delta T_c]}+b,
\end{equation}
where $a$, $b$, $T_c$ and $\Delta T_c$ are fitting parameters, we find that the critical temperature of the thin film layer and the width of the phase transition are $T_c=70.0(2)$ K and $\Delta T_c=0.5(1)$ K, respectively\bibnotesetup{
note-name = ,
use-sort-key = false
}
\bibnote{Immediately after growth this layer had $T_c=82$ K. The subsequent reduction of the critical temperature is attributed to three factors. First, degradation of the sample may have occurred as it was held for a year at ambient conditions. During this time it was mounted and removed from the cryostat several times and came in contact with vacuum grease and vacuum epoxy that may have caused degradation. Second, the thermal contact between the cryostat's cold finger and the superconducting sample may have been less than ideal, although we took measures to maximize it. We pressed the superconducting sample against the cold finger, using a piece of Teflon. Since the temperature sensor is located inside the cold finger, its reading may be lower than the actual temperature of the sample. This is so especially since the sample may have been heated by the green excitation laser. Finally, another source of error is heating of the sample by the proximal microwave wires. Indeed, when microwave power is turned on, we see instabilities in the temperature reading even though the temperature is locked with a proportional-integral-differential (PID) loop. By adjusting the PID parameters, we were able to keep these fluctuations below 1 K.}.


Next, we monitored the sample's response to an external perpendicular magnetic field while the field was increased from zero.

It is important to note that in a thin-film geometry with an external field applied perpendicularly, as in our case, the result of the measurement strongly depends on the location of the probe with respect to the sample. In the case where the superconductor cross-section (perpendicular to the field) is small (relative to the sample thickness), one expects the field lines to be hardly perturbed at the surface, and consequently, when the external field reaches the critical value of $H_{c1}$, vortices will form on the surface of the sample with a homogeneous distribution. However, in the geometry where the field is applied perpendicularly to a thin-film, the field lines concentrate near the edges of the sample due to their expulsion from the superconductor. Hence, the vortices first form at the edges, and as we increase the field they start filling up the area towards the center of the film\cite{ZEL1994}.

The width of the ``vortex free'' area in the center of the sample is given by \cite{ZEL1994}
\begin{equation}
a=\frac{W}{\cosh{\left(H_0/H_f\right)}},
\label{eq:azone}
\end{equation}
where $W$ is the width of the film, $H_0$ is the external applied field, and $H_f=4dJ_c/c$ is the characteristic field for the film geometry (when a significant part of the surface experiences vortices), with $J_c$, $d$, and $c$ being the critical current density, film thickness and the speed of light, respectively.

From this we conclude that near the center of the superconducting square, an extremely high external field is required in order to observe vortices (according to Eq. \eqref{eq:azone}, for $a\rightarrow 0$ we need $H_0 \gg H_f$), while at the edges they appear at rather low fields. We chose to put the probe at a point which is approximately midway between an edge and the center, so the change in the local magnetization (or the measured magnetic field) due to the penetration of vortices may be observed.

Gluing the diamond on the sample with a cryogenic compatible varnish, as was done here, resulted in a distance between the surfaces of the diamond and the sample which is $> 10~\mu$m (measured using a digital gauge). At such a height above the sample, the field measured is the averaged local field of the vortices, namely, $B =\phi_0n$, where $\phi_0=20.7$ G$\cdot\mu$m$^{2}$ is the flux quantum \cite{SCH1997}, and $n$ is the number of vortices per unit area. As mentioned above, in the future we intend to use the setup for imaging vortices. For such a measurement the distance between the detector and the superconductor sample must be kept smaller than $\sim1~\mu$m \cite{BEN1999}. Such proximity is enabled only  by various experimental techniques which we are now exploring. We elaborate on these techniques in Sec. (\ref{onchip}).

Finally, in the experiment, we increase the current in the coil, starting from zero and in increments of 1 A (corresponding to $\sim 4.5\ $G), and record the ODMR spectrum each time. After we cross the penetration field~$H_P$ (the magnitude of the applied field resulting in the onset of vortex formation), we gradually decrease the current to zero. The temperature is kept at 65 K for the whole sequence.

\begin{figure}[h]
\centering
\includegraphics[scale=0.3]{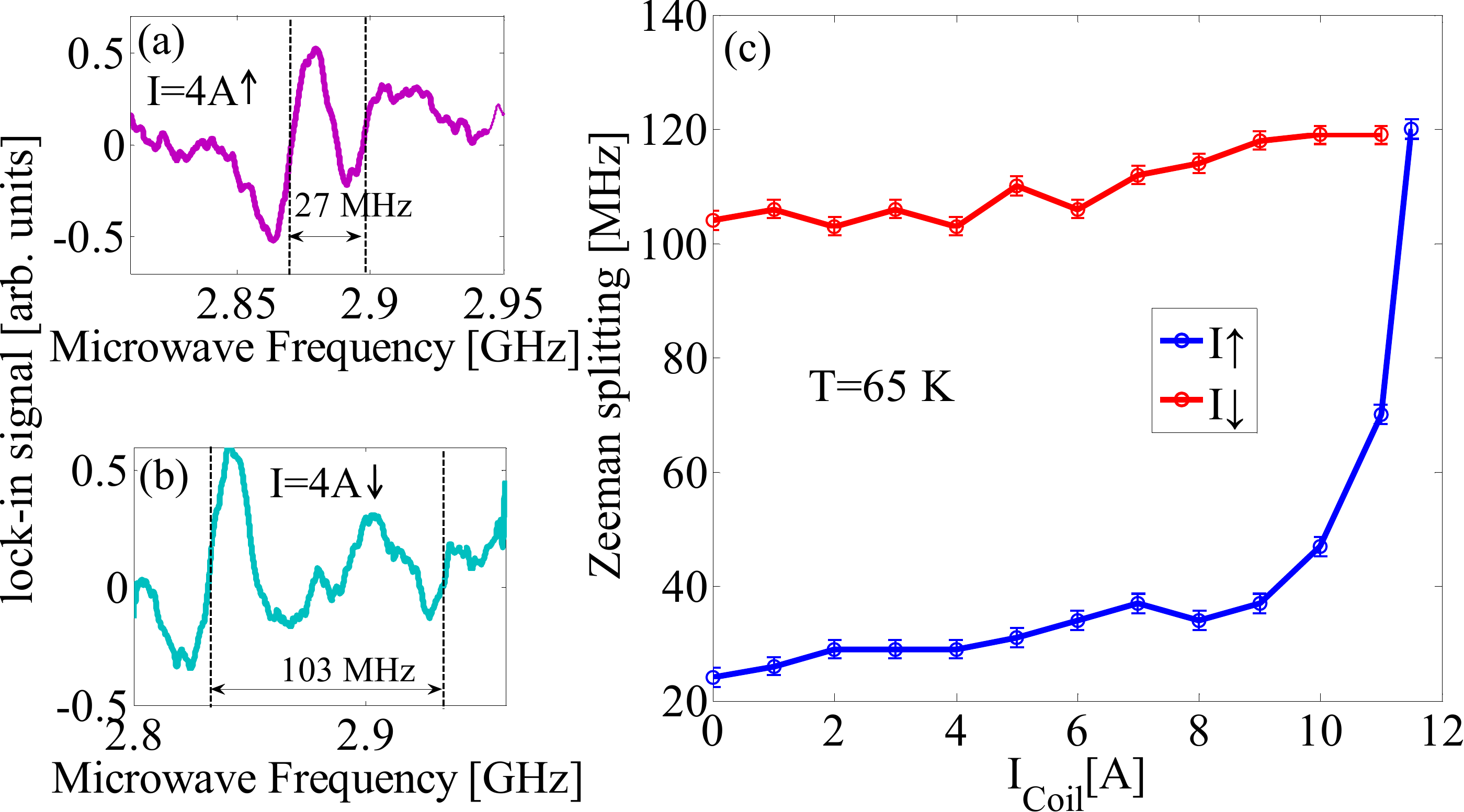}
\caption{(a-b) ODMR spectra taken during the experimental sequence for finding the local penetration field $H_P$. The two different signals, taken with the same external field, reflect different responses of the material: (a) Taken after increasing the field from zero, meaning the system at this spatial point is in the Meissner state, and the external field is screened. (b) Recorded after lowering the field from $\sim 50$ G, which is above $H_P$, thus measuring the field of the pinned vortices. (c)  The Zeeman splitting detected with the NV-diamond sensor during the measurement sequence. The applied magnetic field is proportional to the coil current. The plot demonstrates the local transition from the Meissner state to an intermediate state wherein vortices are in the sample. The blue points correspond to increasing coil current while the red points correspond to the decreasing-current sequence. The blue and the red lines are to guide the eye.}
\label{fig:Hc}
\end{figure}

The results of these measurements are presented in Fig. \ref{fig:Hc}. In Fig. \ref{fig:Hc}(a) we see that below~$H_P$, the Zeeman splitting is relatively small, indicating the absence of vortices. At $I_{coil}\approx10$ A, we observe a sudden increase of the measured Zeeman splitting due to the formation of vortices. Next, the field is gradually eliminated, but a substantial Zeeman splitting of $\sim 103$ MHz is still measured between the two magnetic resonances of the NV centers with $35^\circ$ axes [see Fig. 4(b)]. In the absence of a superconductor, a field of $\sim22.5$ G would normally be required to produce such a splitting. The reason for this hysteresis is pinning of vortices by defects in the lattice: upon crossing the critical field, vortices are generated and trapped by microscopic defects within the sample \cite{SCH1997}. As we lower the field, the vortices remain trapped because the pinning force is field-independent. In this regime we measure the field of trapped vortices.
Figure 4(c) summarizes the measurement sequence. The blue data points were recorded while increasing the current, and the red ones while decreasing it.

We may estimate the penetration field $H_P$ using the data of Fig.~\ref{fig:Hc}(c). The sharp increase in the Zeeman splitting around $I_{coil}$=10~A is interpreted as the formation of vortices at the position of the sensor, which leads to a non-zero magnetic induction $B$ inside the material. The magnetic field corresponding to $I_{coil}=10$ A is found  by measuring the Zeeman splitting induced by this current above $T_c$, wherein the external field is no longer screened. Converting to units of magnetic field we get $H_P=46.2\pm 3.9$~G.

As reflected by the red data points in Fig.~\ref{fig:Hc}(c), the induced magnetic field is almost completely preserved due to flux pinning when the external field is decreased. This trapped flux corresponds to a magnetic field of $B=22.4\pm 0.5$~G.

In addition, we performed an experiment where we cooled the system in a weak field of $\sim 10$ G, down to 40 K. The resulting trapped flux corresponded to a magnetic field of $9.5\pm 0.2$ G, as calculated from the measured Zeeman splitting. This field corresponds to a vortex density of $n=B/\phi_0\approx0.45\pm0.01~\mu$m$^{-2}$. Here, as we cool the system with an applied field the density of vortices will be approximately uniform across the sample.  Under these conditions, and by positioning the sensor less than $1~\mu$m above the surface, it should be possible to image an isolated vortex. Let us note that as our resolution increases, namely, as our pixel size and therefore our detection volume is decreased, the sensitivity will only be suppressed as one over the square root of the volume. In case the sensitivity requires compensation, the thickness of the NV layer may be increased (e.g. by an order of magnitude), and in addition, the density of NV centers may be enlarged.

The values measured in this work are listed in Table I below.

\begin{table}[h]
\centering
\begin{tabular}{| >{\centering\arraybackslash}m{1.3cm} | >{\centering\arraybackslash}m{2cm} | >{\centering\arraybackslash}m{2cm} | >{\centering\arraybackslash}m{2.7cm} |}
\hline
Property & Measurement & Ref. Value & Comments \\
\hline
$T_c$ & $70.0\pm 0.2$ K & $82.2 \pm 0.3$ K & Ref. value measured after sample growth \\
\hline
$\Delta T_c$ & $0.5\pm 0.1$ K & $0.14\pm 0.01$ K & same as above \\
\hline
$H_P$ & $46.2\pm 3.9$ G & None & local measurement \\
\hline
$n$ & $0.45\pm 0.01$ $\mu$m$^{-2}$ & None & --------- \\
\hline
\end{tabular}
\caption{Superconductor-layer properties measured by the NV sensor. The reference values were also measured in our lab. We have used a miniature coil located on one side of the sample to transmit an AC signal of a $\sim1$ kHz frequency. A pickup coil, identical in specifications to the transmitting coil, is placed on
the other side of the sample. The whole structure is shielded with copper rods and dipped in liquid nitrogen. The shielding decreases the rate of sample cooling, enabling us to monitor the pickup coil signal vs. the temperature, in order to obtain
the phase transition curve. For a discussion regarding the difference in the phase transition temperature between the diamond measurement and the reference value, please see Ref.$^{51}$}
\label{tab:sc}
\end{table}



\section{On-chip vortex imaging}
\label{onchip}
This work may  be extended to image vortices.  The first part of this section describes the fabrication of a chip for this purpose.  The second part describes the difficulties and possible remedies concerning the sample-diamond distance.

\begin{figure}
\centering
\includegraphics[scale=0.3]{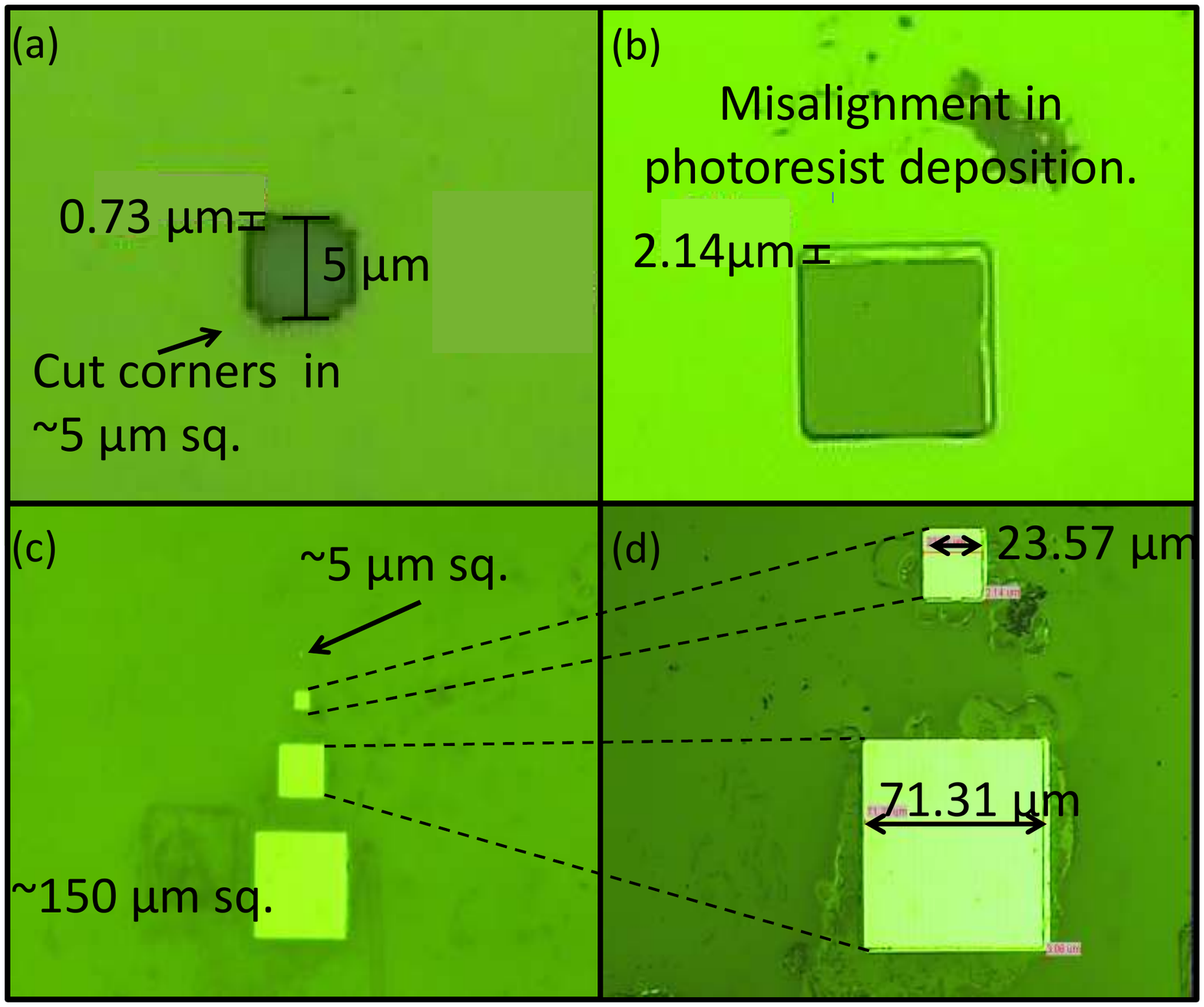}
\caption{Different stages in the fabrication of the superconductor chip. (a) The smallest square on the chip after the first photoresist deposition. To realize a pattern on the MgO substrate, we used a positive mask and etched the YBCO with phosphoric acid. The cut corners are an artifact of the exposure process. (b) Following the YBCO etching, we covered the chip with a photoresist, after writing a negative mask. In this image, of one of the squares, the surface is covered with photoresist except for the desired pattern. A minor misalignment, of $\sim2~\mu$m can be observed. (c) Image of the final chip, after covering it with silver, and lifting off the photoresist. The small square is marked with an arrow as it is not visible on this scale. (d) Zoomed image of two of the squares. The actual sizes, measured by microscope, proved to be somewhat smaller than the planned ones ($25~\mu$m and $75~\mu$m).}
\label{fig:chip}
\end{figure}

We have chosen a pattern of four YBCO squares with varying sizes ranging from $5\,\mu$m to $150\,\mu$m. This pattern will generate boundary conditions on the  magnetic field of the vortices, which we will map using our magnetometer. The small square may also enable us to capture a single vortex. The design is shown in Fig.~\ref{fig:chip}.

Square masks were written on the YBCO layer using a mask-aligner. A photoresist was then deposited onto the sample. This technique is usually referred to as positive mask deposition. Next, we exposed the chip to light and wet-etched it using phosphoric acid at a concentration of $8\%$ for $60$ seconds (etching rate of $5$ nm/sec). We then deposited on the chip a negative mask by writing the same pattern of squares used previously  and subsequently applying the photoresist. At this point, the photoresist is found everywhere except on the squares. After exposure of the resist, we deposited a $100$ nm layer of silver on the chip, using an electron gun. This layer protects the YBCO layer and reflects the green light used for NV sensing to prevent heating of the superconducting thin film. Lifting off the photoresist, we obtain the desired pattern, as shown in Figs.~\ref{fig:chip}(c-d).

\vspace{0.8cm}

Let us now discuss the issue of the sample-diamond distance. As noted, this gap may not be larger than the required spatial resolution. The required resolution depends on the density of vortices if one wishes to image vortex lattices or dynamics, or on the diameter of the vortex if one wishes to image the vortex structure. The required distance may then vary from a few $\mu$m to a few nm.

Generating a small sample-diamond distance is rather challenging. Even if the two surfaces are polished with a surface roughness of a few nanometers, bending due to internal stress, microscopic particles (dirt) and different thermal expansion coefficients, may cause this distance to be too large and nonuniform across the sample. Naturally, the direct deposition of the superconducting material on the diamond ensures a minimal sample-diamond distance. However, cuprate high $T_c$ superconductors such as YBCO have to be epitaxially grown on the substrate, and consequently only substrates that can be lattice-matched to the superconductor (e.g. MgO) might be used. To the best of our knowledge, diamond is not one of these substrates, and thus a buffer layer of several tens of nm is usually used \cite{CUI1992}. As a result, YBCO growth on diamond is a rather complicated process which requires comprehensive research. As a first step, one may consider to deposit a superconductor of the type Nb on a diamond. However, Nb exhibits a relatively low $T_c$ ($\sim9.2$ K in the bulk \cite{FIN1966} and even lower for thin films), meaning a significant effort must be invested in the cooling of the sample while it is in thermal contact with a diamond which is being heated by light and MW radiation.

Another option to consider is the use of thin diamond slabs ($10\rm\,\mu$m or even less) which are expected to be more flexible and thereby follow the SC layer, even if both surfaces suffer from initial bends, or local topography fluctuations. The thinning of samples (having for example an initial thickness of $\sim30\,\rm \mu$m) might be performed using the deep reactive ion etching technique (DRIE).

A third option is the welding of the diamond to the YBCO layer by depositing nanometer-sized gold dots on both surfaces and heating them while pressing one against the other. This method is usually referred to as  thermocompression bonding. Since the diamonds are much smaller than the conventional substrate normally used in chip lithography (the size of the diamond is, for example, $1\times1\times 0.08$ mm while conventional wafers are of $50$ mm diameter and of $0.5$ mm thickness), such a process is not trivial, and has to be carefully developed and executed.

\section{Summary and outlook}
\label{Summary}
We have presented magnetic-field measurements above a thin-layer superconductor using a diamond magnetometer. Using a $10$ nm layer of NV centers formed via nitrogen implantation and annealing, we measured the superconducting phase transition, as well as the local vortex penetration field. We have observed vortex pinning, and furthermore, determined the surface density of vortices in the layer following the cooling of the sample in a $10$ G field. For future work, we constructed a superconductor chip which is suitable for vortex mapping.

Assuming the density of vortices measured here, standard optical resolution ($\sim 400$ nm in our case) will be sufficient to map the distribution of vortices in the sample. Comparing to our current setup this task will demand a short sample-diamond distance ($< 1\ \mu$m) and a detection setup based on a camera or on a high-bandwidth scanning system. These two tasks are the subject of ongoing experimental efforts.

To image the core of a vortex we would require sub-diffraction imaging methods.  Such sub-diffraction imaging methods have previously been used with NV centers (see Ref.~\cite{HAN2009}, for example). Hence, imaging of vortex patterns and cores with nanoscale resolution and better than $1\ \mu$T sensitivity should be possible with this technology.


We thank the members of the Atom Chip group at Ben-Gurion university and the team of the fabrication facility. We are specifically appreciative of the assistance from Menachem Givon, Yaniv Bar-Haim, and Yacov Berenstein. We thank V. Z. Kresin and Eli Zeldov for stimulating discussions. This work was supported by the NATO Science for Peace program (SfP 983932), the AFOSR/DARPA QuASAR program, a starter grant award from the Spectroscopy Society of Pittsburgh (LSB), a Beckman Young Investigator Award (LSB) and NSF Grant No. ECCS-1202258. Dmitry Budker and Ron Folman gratefully acknowledge support by the Miller Institute for Basic Research in Science, University of California, Berkeley.


\bibliographystyle{apsrev4-1} 
\bibliography{referencessc4}     

\begin{thebibliography}{57}
\bibitem [{1()}]{1}%
  \BibitemOpen
  \href@noop {} {}\bibinfo {note} {Immediately after growth this layer had
  $T_c=82$ K. The subsequent reduction of the critical temperature is
  attributed to three factors. First, degradation of the sample may have
  occurred as it was held for a year at ambient conditions. During this time it
  was mounted and removed from the cryostat several times and came in contact
  with vacuum grease and vacuum epoxy that may have caused degradation. Second,
  the thermal contact between the cryostat's cold finger and the
  superconducting sample may have been less than ideal, although we took
  measures to maximize it. We pressed the superconducting sample against the
  cold finger, using a piece of Teflon. Since the temperature sensor is located
  inside the cold finger, its reading may be lower than the actual temperature
  of the sample. This is so especially since the sample may have been heated by
  the green excitation laser. Finally, another source of error is heating of
  the sample by the proximal microwave wires. Indeed, when microwave power is
  turned on, we see instabilities in the temperature reading even though the
  temperature is locked with a proportional-integral-differential (PID) loop.
  By adjusting the PID parameters, we were able to keep these fluctuations
  below 1 K.}\BibitemShut {Stop}%
\end{thebibliography}

\end{document}